\documentclass[prl,twocolumn,final,superscriptaddress,showpacs]{revtex4}
\usepackage{graphicx}
\newcommand{\bra}[1]{\langle {#1} |}  
\newcommand{\ket}[1]{| {#1} \rangle}

\begin{document}

\title{Electronic structure of overstretched DNA}

\author{Paul~Maragakis}
\affiliation{Physics Department and Division of Engineering and
  Applied Sciences, Harvard University, 02138 Cambridge MA}
\author{Ryan Lee Barnett}
\affiliation{Physics Department and Division of Engineering and
  Applied Sciences, Harvard University, 02138 Cambridge MA}
\author{Efthimios Kaxiras}
\affiliation{Physics Department and Division of Engineering and
  Applied Sciences, Harvard University, 02138 Cambridge MA}
\author{Marcus Elstner}
\affiliation{University of Paderborn, Fachbereich Physics,
D-33098 Paderborn, Germany.}
\affiliation{German Cancer Research Center, Department of Molecular
Biophysics, D-69120 Heidelberg, Germany.} 
\author{Thomas~Frauenheim}
\affiliation{University of Paderborn, Fachbereich Physics,
D-33098 Paderborn, Germany.}

\begin{abstract}
  Minuscule molecular forces can transform DNA into a structure that
  is elongated by more than half its original length.  We demonstrate
  that this pronounced conformational transition is of relevance to
  ongoing experimental and theoretical efforts to characterize the
  conducting properties of DNA wires.  We present quantum mechanical
  calculations for acidic, dry, poly(CG)$\cdot$poly(CG) DNA which has
  undergone elongation of up to 90 \% relative to its natural length,
  along with a method for visualizing the effects of stretching on the
  electronic eigenstates.  We find that overstretching leads to a
  drastic drop of the hopping matrix elements between localized
  occupied electronic states suggesting a dramatic decrease in the
  conductivity through holes.
\end{abstract}


\maketitle

%Biological macromolecules like DNA are thought to hold great promise
%as components of future nanoscale devices.  The highly desirable
%properties of self-assembly in combination with the molecular
%factories of enzymes that are able to multiply, cut, and join DNA are
%indications that this promise may one day be fulfilled.  In this
%respect, understanding the nature of conductivity in DNA and more
%generally its electronic properties, is of paramount importance.  
The possibility of efficient charge transfer along the stacked $\pi$
orbitals in DNA was proposed 40 years ago~\cite{ELEY1961}.  In recent
years its conducting properties were tested in several contexts: DNA
has been used as a template on top of which nanowires are
built~\cite{BRAUN1998} and there is hope that self-assembled DNA
junctions between conductors will function soon.  There has also been
renewed interest in building and characterizing molecular wires of
plain DNA between two metal
leads~\cite{FINK1999,PORATH2000,DEPABLO2000,KASUMOV2001,YOO2001},
possibly combined with changes of the major constituents of the
molecule to improve its properties~\cite{RAKITIN2001}.  There is
ongoing debate as to whether DNA is conducting, semi-conducting,
insulating or all of the above~\cite{FINK2001}.  There are several
reasons for this controversy. There is great diversity of the DNA forms in terms of
its composition, length, and structure.  Counterions and impurities
that can attach to the phosphate groups and the grooves of the
molecule need to be taken into account.  Finally,
environmental factors such as temperature and contact resistance can
play a major role.

%The considerable progress made in the past
%few years needs to be complemented by improved understanding at the
%microscopic level.  Biochemical methods mostly deal with ensemble
%measurements performed on billions of molecules in solution, whereas
%characterization of device components by necessity must deal with each
%individual element.  The methods for manipulating DNA at the single
%molecule level are not as evolved as the ensemble methods, but
%progress has been made both in terms of imaging (electron
%holography~\cite{FINK1999}, and
%microscopy~\cite{HANSMA1992,WOOLLEY2000}) and controlling individual
%molecules (controlled hydrodynamic flow~\cite{SMITH1992}, manipulation
%with magnetic or optical tweezers~\cite{SMITH1996}, detection through
%nanopores~\cite{LI2001}).

Experiments done 50 years ago by Wilkins et al.~\cite{WILKINS1951},
suggested that
overstretched DNA (that is, substantially longer than its natural
length) undergoes a transition to a structure that can accommodate
elongation up to twice the length of relaxed DNA.  A major
breakthrough in understanding the mechanical properties of DNA was
achieved through single molecule stretching
experiments~\cite{SMITH1996,CLUZEL1996,STRICK2000}.  It was shown that
by gradually increasing the axial stretching force $F$ along the DNA
helix, the molecule first uncoils, then, at $F~$ 5 pN, it reaches its
natural length, and then exhibits a stiff elastic response for forces
of up to 50 pN.  At $F~$ 50 to 80 pN, DNA
undergoes a pronounced and abrupt structural transformation to a yet
unknown structure that is elongated by more than 50\%.  Further
increase in force leads to breaking after it has stretched to twice
its natural length.  In this letter we argue that these overstretched
forms of DNA are relevant in the quest for novel device components, we
develop a visualization method for understanding the nature of
electronic states in these forms, and we present the first detailed
discussion of their electronic properties.

We use an efficient quantum mechanical electronic
structure method~\cite{ELSTNER1998a} that
treats all the valence electrons and that
accounts for the charge transfer involved in biological
molecules and semiconducting surfaces. This approach, called
self-consistent-charge density-functional tight-binding method, is
based on an expansion of the Kohn-Sham energy
functional~\cite{HOHENBERG1964,KOHN1965} in terms of the charge
density.  The zeroth order term results in a tight-binding
Hamiltonian, and the second order term incorporates the charge
redistribution.  The method is able to deal with the very large
numbers of atoms involved in biologically relevant systems.  As a
test, we performed calculations on a periodic homopolymer of
poly(C)$\cdot$poly(G) with 10 base pairs per turn and reproduced the
results of dePablo {\em et al.}~\cite{DEPABLO2000}, which to our
knowledge is the only {\em ab initio} electronic structure calculation
of a periodic DNA structure~\cite{LADIK}.  Our results are in excellent agreement
to their calculations: we find a 2.1 eV bandgap between the
highest occupied and lowest unoccupied eigenstates, compared to 2.0 eV
reported in Ref.~[\onlinecite{DEPABLO2000}].  The wavefunctions of the
highest occupied valence band are located on the guanines while the
wavefunctions of the lowest unoccupied band are located on the
cytosines, also in agreement with Ref.~[\onlinecite{DEPABLO2000}].

\begin{figure}[!htp]
  \includegraphics[width=0.97\columnwidth]{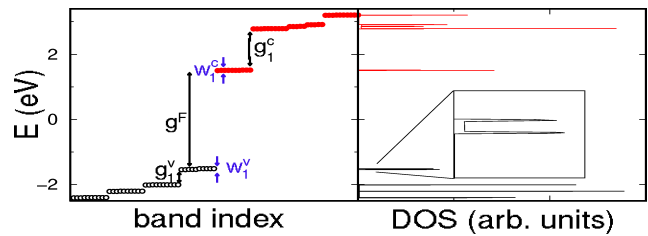}
  \caption{The band structure around the Fermi energy
    (left) with the corresponding density of states (right) for the
    periodic poly(CG)$\cdot$poly(CG) form.  The inset shows a detail of the density of
    states at the highest occupied valence band.  We denote the
    bandgap around the Fermi energy (g$^F$), the widths of the highest
    valence (w$^v_1$) and lowest conduction (w$^c_1$) minibands, and
    the gaps between these and the neighboring minibands (g$^v_1$ and
    g$^c_1$).}
  \label{fig:bands}
\end{figure}

For the purpose of the present study, calculations were performed on a
variety of DNA polymers with different compositions employing both
infinite periodic and finite structures.  We limit our discussion to
poly(CG)$\cdot$poly(CG) as a representative structure for which well
established overstretched configurations have been
published~\cite{LEBRUN1996}.  Our results on finite DNA polymers show
the existence of states localized at the opposite ends of the
molecule.  The energy levels of these states lie between the regular
bands that appear in the periodic structures.  We removed these states
before analyzing the electronic properties, and will refer to the
remaining states as the ``restricted band structure'' of the
non-periodic configuration. As a test, we compared 18 base pair
restricted band sturcture with periodic structures with 10 base pairs
per unit cell and found the band structures virtually
indistinguishable.  A boundary region of 3 base-pair layers at each
end of the molecule was sufficient to contain the spurious end states.
Fig.~\ref{fig:bands} displays the band structure of a periodic
sequence with a unit cell consisting of 10 base pairs.  The sharp
features of the density of states are due to the one dimensional
nature of DNA.  In all periodic sequences there is an extra symmetry,
which for the poly(CG)$\cdot$poly(CG) structure is the translation by
two base pairs along the stack, together with a rotation of 72 degrees
around the helical axis: this symmetry is generated by the screw
operator that is isomorphic to a pure translation operator.
Structures obtained by short (50 ps) molecular dynamics runs at 50 K
using classical force fields~\cite{CORNELL1995}, indicate that minor
deformations of the ideal configuration destroy the sharp features in
the electronic structure.

We now move on to a discussion of overstretched DNA.  First we
elaborate on how this is relevant in the construction of DNA
nanowires.  One way to attach DNA molecules
to substrates is the technique of ``molecular
combing''~\cite{BENSIMON1994}.  This method uses the meniscus forces
that develop between the solution surface and the device surface
template to uncoil the DNA and stretch it between two electrodes as
the device is pulled out of the solution.  A
description of recent experiments estimates that the forces that are
exerted on the DNA molecules are of order 100 pN~\cite{KLEIN2001},
more than enough to stretch them past their natural length.  This is
indicated by recently built devices~\cite{WU2001} where the distance
between the electrodes was about 30\% longer than the natural length
of the DNA molecules.  The overstretching can
have a variety of consequences on the electronic states.

Elongated DNA structures have been determined in the
pioneering study of Lebrun and Lavery~\cite{LEBRUN1996}, which model
the adiabatic elongation of selected DNA molecules.  There are two
distinct modes of overstretching corresponding to pulling the 
opposite 3'-3' ends and
5'-5' ends.  The former mode leads to DNA
unwinding, consistent with a recent estimate of the helicity of
overstretched DNA~\cite{LEGER1999}, whereas the latter leads to a
contraction of the diameter of the helix.  Both overstretching modes
can accommodate elongations of up to 90\% without breaking the
molecule. 
%~\cite{acidic}.

%~\footnote{We obtain the structure of phosphoric acid by
%  energy minimization and use its local geometry (lengths, angles and
%  dihedral angles) for the proton binding to the phosphate group.  The
%  lowest energy configuration corresponds to the proton attaching to
%  the oxygen which points towards the minor groove.}.

We next present the inherent electronic states of the overstretched
structures and then consider electron transport aspects.  
For all the calculations we used acidic, dry, 18 base-pair strings of
DNA, and we performed the projection of the end states in the manner
stated before.  The quasi one-dimensional structure of all the forms
leads to the formation of clear minibands in the energy spectrum,
similar to those shown in Fig.~\ref{fig:bands}.  The band structure of
the overstretched forms have an overall resemblance to the band
structure of the original unstretched form, and many of the minibands
near the Fermi energy can be continuously mapped to their counterpart
in the unstretched configuration.  This is mostly because of the
molecular structure of the overstretched forms as will become clear in
the following discussion.  For some of the most deformed structures,
especially the 90\% overstretching in the 5'-5' mode, several of the
minibands are mixed and small but clear gaps present in less deformed
structures disappear.

\begin{figure}[!htp]
   \includegraphics[width=0.97\columnwidth]{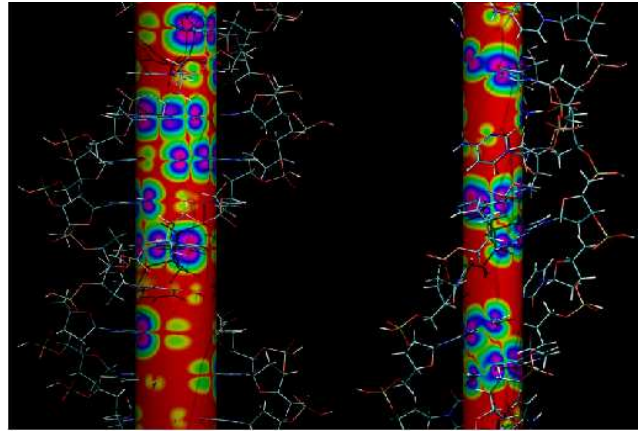}
   \caption{The left panel shows
     a piece of poly(CG)$\cdot$poly(CG) B-DNA.  Along its helical axis
     is a cylinder on which we plot the contours of the valence band
     electron density.  The right panel shows the same
     region for a 30 \% stretched structure.}
   \label{fig:coverpage}
\end{figure}

We have devised a new scheme for visualizing the electronic states of
DNA, which naturally accounts for the symmetry of this molecule.  The
basic concept of this scheme is to unfold contour plots of the band
electron density evaluated at cylindrical surfaces centered on the
helical axis, as shown in Fig.~\ref{fig:coverpage}.  In this figure we
show the structure at its natural length together with one that is
elongated by 30\% in the 3'-3' overstretching mode.  With the help of
these plots we can understand most of the band structure features of
the stretched forms.  The elongation to the overstretched form is
achieved by changing the dihedral angle configuration of the backbone
of the molecule, thus leaving the local part of the orbitals
essentially intact. 

\begin{figure*}[tpbh]
  {  \rotatebox{-90}{\includegraphics[width=0.18 \textwidth]{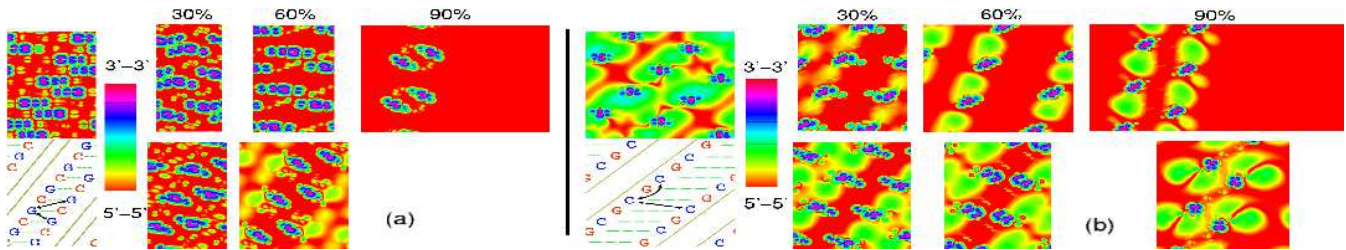}} }
  \caption{Cylindrical contour map of the highest valence (a) and
    the lowest conduction (b) band electron density, for the
    poly(CG)$\cdot$poly(CG) DNA (top left) and the 30\%, 60\%, and
    90\% overstretched forms (right).  The bottom left panel of each
    figure shows a drawing of the CG base pairs along the cylinder; the
    arrows indicate bases with the largest couplings.  In Fig.~(a) the
    90 \% overstretched 5'-5' form is not shown because the highest
    valence band mixes with the lower bands.  The vertical axis covers
    a fixed region corresponding to 10 base pairs in the unstretched
    form.  The horizontal axis runs from 0 to $2 \pi$.  The radius of
    the cylinder follows the guanine C5 atom in (a), and the cytosine
    C6 in Fig.~(b).  The cylinder surface unit element is kept fixed:
    this results in a variation of the length of the horizontal axis,
    proportional to the cylinder radius.  The color coding (shown in
    the color bar between the unstretched and stretched graphs) is
    fixed  
    between different structures and covers 5 orders of magnitude.  }
  \label{fig:stm}
\end{figure*}

In Fig.~\ref{fig:stm}~(a) we present the valence band electron density of
selected stretched poly(CG)$\cdot$poly(CG) structures using our
visualization scheme.  We choose 
the radii of the cylinders to follow the atom with the largest
electron density for the band that we present.
%The radius
%of each cylinder was modified to follow a given atom (in this case C5
%of guanine) and thus varies between the different structures.  The
%length along the cylindrical axis is kept fixed, equal to the natural
%length of the 10 base-pair structure, and the angle is scanned over
%the full $2 \pi$ range.  Scientific
In the unstretched structure, the $\pi$-band is along the guanines.
Upon elongation, orbitals from the picture frame are removed while the
remaining ones rotate partially as dictated by the new orientation of
base pairs.  The poly(CG)$\cdot$poly(CG) structure contains
pairs of guanines that are closer to each other.  These guanines
display an overlap of the electronic densities as seen in the top 
left frame of Fig.~\ref{fig:stm}~(b).  Elongation in the 3'-3' mode
keeps such paired guanines close to each other which is
reflected in the overlap in their electronic densities; the overlap
of the electronic densities between guanines from different pairs
decreases with elongation. 
%\begin{figure*}[htp]
%%  \includegraphics[width=0.92\columnwidth]{conductiondens.png}
%  \includegraphics[width=0.98\columnwidth]{conduction-horiz-dens.png}
%  \caption{The highest conduction band electron density with the same
%    conventions as Fig.~\ref{fig:stm}.  The larger cylinder radii are
%    related to the position of C6 of cytosine which has the maximum
%    electron density.  The arrows in the bottom left panel indicate
%    the coupling between the closest neighbor and the second neighbor
%    cytosines.}
%  \label{fig:stm:conduction}
%\end{figure*}
In Fig.~\ref{fig:stm}~(b) we show the conduction band electron
density.
%This time we select the radius of the cylinder so that we
%follow atom C6 of cytosine (where most of the electron density of the
%lowest conduction band is localized).  
Note again how the orbitals
rotate as the structure is being overstretched, following the rotation
of the bases.  Note that in the extreme 90\% stretching case, the
orbitals become perpendicular to the helical axis.

\begin{figure}[!htp]
  {\rotatebox{-90}{\includegraphics[width=0.31 \columnwidth]{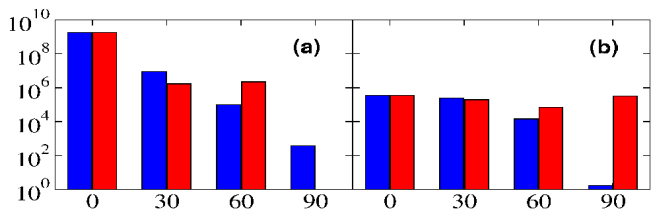}} }
  \caption{The values of \mbox{$|t_{ij}|^2$} from Eq.~(1) for (a) the
    valence and (b) the conduction band.  The values are given eV$^2$
    multiplied by a factor of 10$^{12}$ to facilitate relative
    comparisons, for overstretching of 0, 30, 60, 90 \% relative to
    the natural length.  The blue (left) column at each entry
    corresponds to the 3'-3' stretching mode and the red (right)
    column to the 5'-5' mode.  In Fig.~(a), the entry at the
    90 \% stretching of the 5'-5' mode is missing because of the mixing the
    valence band with lower bands.}
  \label{fig:t2}
\end{figure}

The changes in the structural features should be of profound
consequence to any model dealing with electron transfer since, to a
first approximation, the overlap matrix elements depend exponentially
on the distance between orbitals.  Because of the $\pi$ character of
the states near the Fermi level, a non-parallel alignment will
strongly affect the values of the overlap integrals.  In order to
quantify these effects, we considered the nearly degenerate
eigenstates in the first valence miniband.  We used the linear
combination of these states that gives normalized, orthogonal states
$\ket{\Psi_i}$, which are maximally localized along the helical axis
within the subspace spanned by the delocalized states.
%~\cite{transform}.
%\footnote{The
%  transformation was performed with the unitary matrix which
%  diagonalizes the position operator along the helical axis in the
%  delocalized eigenstate basis through a similarity transformation.}.
The resulting valence states are localized on guanine bases.  With
these states, we calculated the hopping matrix elements:
\begin{equation}
\label{eq:tij}
\mbox{$t_{ij}=-\frac{1}{2}\bra{\Psi_i}\nabla^2\ket{\Psi_j}$},
\end{equation}
between nearest-neighbor sites.  Note that for the
poly(CG)$\cdot$poly(CG) structure every guanine has 2 guanine
neighbors, one above and one below as indicated in Fig.~\ref{fig:stm}, 
separated by distances of 5.3 \AA\ and 4.2
\AA.
The larger of the two matrix elements corresponds to hopping between
sites that are close together; its value remains roughly constant for
the 30$\%$ overstretched structures.  The smaller of the two matrix
elements, which corresponds to sites that are further apart, is
dramatically reduced for all the overstretched forms.  Since the two
different hopping terms are connected in a 1-D series the smallest
term will determine the bottleneck for electron transport.  The
absolute square values of these small hopping matrix elements between
these neighbors are given in Fig.~\ref{fig:t2}~(a).  We infer from
this figure that if the conduction mechanism is through holes there
should be a dramatic drop in conductivity with 
30\% 
overstretching, which in the 3'-3' mode drops further with
elongation.  
In
Fig.~\ref{fig:t2}~(b) we give the hopping matrix elements between
states in the conduction band.  In the cases where the pairing between
nearest neighbor pairs have very different matrix elements we only
list the one that would be the bottleneck for this channel.  
Due to
the stronger localization of the conduction band and because of the
rather large distance between the occupied sites, the hopping matrix
elements are substantially smaller than the ones between the valence
states in the structure of natural length.  In fact, it is the site 
that is two bases away which has the
strongest overlap in this structure: the C6 atoms
between these bases are actually closer to each other (9.0 \AA) than
the C6 atoms between consecutive base pairs (10.0 versus 10.3 \AA).
The first and second neighbor hopping terms are indicated in
Fig.~\ref{fig:stm}~(b).  Since these terms are connected in
parallel, this time it is the maximum of the two that will play the
dominant role in the conduction mechanism and it is this one that is
included in Fig.~\ref{fig:t2}~(b).  We note that if conduction
takes place by electrons the conductivity drops with overstretching in
the 3'-3' mode but not necessarily in the 5'-5' mode.

The concept of overstretched DNA in relation to electronic device
components appears relevant to a variety of experimental and
theoretical issues.  Experimentally it is possible, and sometimes even
unavoidable, to create overstretched DNA molecules.  The window of
allowable forces that can be applied to stretch DNA to its natural
length, but not overstretch, is small (from 5 pN to 50 pN) compared to
the forces applied during molecular combing.  Another important aspect
is owed to the nature of the extreme stretched forms, it seems
possible to expose the bases to the exterior of the molecule allowing
impurities to enter the crucial base-pair stack region; this could
result in controlling the electronic properties with doping.  This
 might also enable direct visualization of parts of the
electronic wavefunctions which was is not possible in the natural DNA
forms because of the insulating backbone layer.  On the theoretical
level, the major existing models (superexchange~\cite{JORTNER1998},
hopping~\cite{JORTNER1998}, and polaron~\cite{HENDERSON1999}), of pure
DNA conductivity will be drastically affected by the consequences of
overstretching, including changes in the hopping matrix elements and
in the low-energy vibrational modes.  For long fibers of homogeneous
DNA (either a homopolymer, or a short repeated sequence) we expect
stretching to appear homogeneously.  Regions of varying richness in AT
or GC mixture would induce different lengthenings at different parts
of the molecule, permitting conduction mechanisms to change along the
molecule. 
%The possibilities seem endless.

We thank Leo Kouwenhoven and Nina Markovic for sharing their
experimental insight with us.  We are grateful to Richard Lavery for
providing the overstretched structures.  We thank Hatem Mehrez for his
critical reading of the manuscript.  This work was supported in part
by Harvard's Materials Research Science and Engineering Center, which
is funded by the National Science Foundation.  This work was partially
supported by National Computational Science Alliance under MCB010005N
and utilized the Boston University Scientific Computing and
Visualization facilities.  RLB was supported by a national science
foundation graduate research fellowship.

\end{document}